# Parameterized Hardware Design with Latency-Abstract Interfaces

Rachit Nigam
MIT CSAIL
Cambridge, USA

Ethan Gabizon*
Cornell University
Ithaca, USA

Edmund Lam*
Cornell University
Ithaca, USA

Carolyn Zech
MIT CSAIL
Cambridge, USA

Jonathan Balkind
University of California
Santa Barbara, USA

Adrian Sampson
Cornell University
Ithaca, USA

## Abstract

Hardware designs must use latency-insensitive (LI) interfaces when timing is input-dependent. When timing is input-independent, designs should use latency-sensitive (LS) interfaces for maximum performance. However, designs commonly use LI interfaces to integrate with externally generated LS modules–from, e.g., IP generators, high-level synthesis, or domain specific languages. In every fully integrated design, such uses of LI represent *pure overhead*. The challenge is that generators can dramatically change timing interfaces of the modules to meet performance objectives, and LI interfaces act as a useful *design abstraction* and enable timing adaptation.

We define *latency-abstract* (LA) interfaces, a new design abstraction, which provide the timing adaptability of LI interfaces at design-time and the efficient integration of LS interfaces. LA interfaces use *output parameters,* a novel compile-time mechanism for child modules to return values parent modules, to abstract and encapsulate timing behaviors at design time. During design elaboration, LA interfaces are compiled into efficient LS interfaces based on parameter values.

While an attractive option, LA interfaces inherit the complexities of parameterized hardware design: the user must reason how parameters influence timing behaviors of modules and ensure that designs adapt to interface changes. To address this challenge and demonstrate the utility of LA interfaces, we design Lilac, a parameterized HDL that uses a type system track the influence of parameters on timing behaviors and formally guarantee that *every parameterization* of an LA design results in a circuit without structural hazards. We demonstrate Lilac's efficacy by using it to implement parameterized designs and integrate designs generated from external tools. We show that LA designs use 26–33% fewer chip resources and achieve 6.8% better maximum frequencies than comparable LI implementations.

*Equal contribution.

## 1 Introduction

Interfaces create the foundation for modular design: they abstract away implementation concerns while providing sufficient details to enable compositional and efficient integration. In hardware design, interfaces consist of both *structural details*, such as the bitwidths of input and output ports, and *temporal behaviors*, such as how often a module can accept new inputs.

Hardware interfaces' temporal behavior broadly falls into two categories: latency sensitive (LS) or latency insensitive (LI).[1] With LI interfaces, these concerns take the form of explicit control signals, such as *valid* and *ready* ports that indicate when a module has produced an output or is ready to accept an input. LS interfaces, on the other hand, specify input-independent timing behaviors ("a four-cycle multiplier that accepts new inputs every other cycle") and are efficient because communicating modules can simply use the already-available clock signal for synchronization. LI interfaces come with fundamental overhead of extra synchronization logic, but they are *necessary* for handling hardware units where latencies are fundamentally input-dependent.

Hardware designers also use LI interfaces to decouple modules from each other, even when those modules have input-independent latency. One particularly important use case is when integrating *generated* hardware modules using core generators [1, 6, 9], high-level synthesis [30, 36], or accelerator design languages [8, 11, 12, 22, 27]. Even when a generated hardware module could technically use an LS interface, it would be awkward to use because the timing behavior depends on the generator's inputs. For example, changing the arithmetic precision of a floating-point core generator might change the resulting module's latency and pipeline depth; the designer would then need to manually inspect the tool's output and carefully cascade the change throughout their design. In this scenario, LI interfaces are a *convenient but expensive* design abstraction: an LS interface would suffice, but it would require frequent and invasive changes whenever the generated module changes.

[1] Though these terms focus on latency, they are used for interfaces that capture *both* latency and initiation interval information.

There is an alternative: using compile-time parameterization constructs to adapt to modules' changing timing behaviors and to automatically emit the correct LS interface. We define a *latency-abstract* (LA) design as the approach that uses parameterized LS designs to abstract over timing details. Because interfaces in LA designs are a compile-time abstraction, they provide the best of both worlds: the efficiency of LS interfaces and the flexibility of LI interfaces.

However, mainstream HDLs often make LA design infeasible for two reasons:

1. Parameters in mainstream HDLs *flow in the wrong direction* (Figure 3). They make it easy to pass values from parent modules to the child modules they instantiate: for example, code that instantiates a multiplier might specify its numerical precision as a parameter. But when using hardware generators, it is the *generator* that determines timing behavior, not the module that uses the generator. There is no straightforward way for the generated multiplier to pass parameters capturing timing behaviors "upward" to its parent. Developers therefore resort to *ad hoc* and error-prone workarounds such as using hierarchical parameter assignments [19] or external configuration files [31].
2. LA interfaces require subtle reasoning about the influence of parameters on timing behaviors. Parameterized designs are already difficult: they express *families of circuits*, requiring testing across many different parameterizations to gain confidence in their correctness. The addition of complex, parameter-dependent timing behaviors make them even more unappealing to use.

Together, these limitations make LA interfaces challenging to use and leave users of mainstream HDLs with an undesirable trade-off: pay the cost of an inefficient LI interface or risk writing brittle parameterization code.

We address the first challenge by introducing *output parameters*, a novel, compile-time construct that enables child modules to return parameter values to their parents. Output parameters precisely capture the kind of inverted flow needed to express the interfaces of generated modules. We show that LA interfaces with output parameters can express generated module interfaces correctly and concisely (§6).

We address the second challenge by designing Lilac, a parameterized HDL that uses output parameters to enable compile-time reasoning and verification of LA circuits. Lilac programs can explicitly capture LA interface and integrate generated modules (§3). Lilac uses a novel type system (§4) based on timeline types [28] to ensure that *every possible parameterization* of a design will correctly interface with such modules and lowers them into efficient circuit descriptions

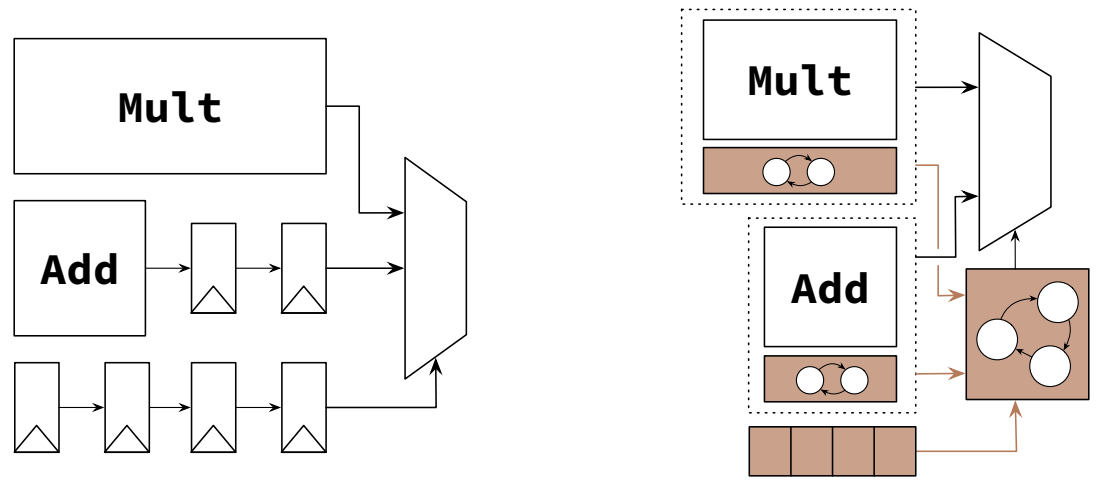

**(a)** LS FPU, fixed timing. **(b)** LI FPU, flexible timing.

**Figure 1.** Circuit diagrams for different arithmetic unit (FPU) implementations that can adapt to changing timing interfaces of multipliers and adders.

```
module FPU(clk, a, b, op, out);
  wire [31:0] add_o, mul_o, add_d[1:0]; wire [3:0] op_d;
  Add(.clk, .a, .b, .out(add_o)) add;
  Mul(.clk, .a, .b, .out(mul_o)) mul;
  always @(posedge clk)
    op_d <= { op, op_d[3:1] };
    add_d[1] <= add_o; add_d[0] <= add_d[1];
  assign out = op_d[0] ? add_d[0] : mul_o;
endmodule
```

**Figure 2.** FPU with four-cycle multiply and two-cycle adder. The implementation delays the adder's output and the op signal to balance the pipeline.

(§5). Finally, we demonstrate that Lilac can be used to implement efficient LA designs which have substantially lower resource usage and higher clock frequencies compared to similar LI designs (§7). These features let Lilac flexibly express LA designs that offer the compositionality and correctness of LI interfaces without their overheads.

## 2 Integrating Generated Hardware

We motivate the need for latency-abstract interfaces by designing a floating-point unit (FPU) that integrates high-quality adder and multiplier implementations generated by FloPoCo [6], an FPGA-focused floating-point core generator. This section examines what integrating such generated modules looks like when using LS, LI, and parameterized design with current HDLs. We then highlight the potential benefits of LA design and the challenges it poses with traditional HDLs.

### 2.1 Latency-Sensitive Interfaces

FloPoCo accepts as input which computation to implement (add, multiply) and performance goals (frequency, FPGA family) to target; it outputs a pipelined, LS implementation and reports its latency. Alternating the performance goals may change the LS interface to the module in unpredictable ways.

Our initial FPU implementation (Figure 1a) uses LS interfaces. We run FloPoCo with specific performance characteristics and use the latency information to integrate the generated modules. Figure 2 overviews the implementation code for a four-cycle multiplier and two-cycle adder: the implementation forwards the inputs directly into the modules and selects the output based on the op signal. Since the multiplier takes two extra cycles, the implementation delays the output from the adder and the op signal to balance the pipeline (lines 5–7).

While straightforward, this implementation has a significant limitation: trying a new design point requires changing the pipeline balancing logic. This is because FloPoCo can change the timing behaviors of the implementations it generates, and our LS implementation fundamentally relies on the specific timing characteristics of the adder and multiplier. This also limits performance portability: changing the FPGA target would require manual changes to the control-logic code.

### 2.2 Latency-Insensitive Interfaces

The fundamental problem with our LS implementation is that it *cannot adapt to changing timing behaviors*. Latency-insensitive (LI) interfaces [5] abstract away timing details using synchronization signals. A common LI interface style is a *ready–valid handshake* that synchronizes one producer and one consumer. The producer uses a 1-bit *valid* signal to indicate that a separate data signal is meaningful; the consumer uses a 1-bit *ready* signal to tell the producer that it is ready to consume the data. When ready and valid are both asserted in the same clock cycle, a transaction occurs.

We can decouple our FPU's control logic from the specific timing behaviors of the compute modules by wrapping them with a ready–valid interface that provides a stable interface. With this approach, whenever FloPoCo generates a new adder or multiplier, we can locally change the LI wrapper to generate the correct ready and valid signals; the FPU control logic can remain unchanged. Figure 1b visualizes this design: the adder and multiplier have extra logic that tracks their ready and valid signals which are then used by the finite state machine (FSM) within the FPU to control the flow of the data. The FPU also has to use a FIFO queue to hold the op signal so it can correctly multiplex between the adder and multiplier outputs. The extra resources used for the coordination logic can quickly dominate the total cost of smaller circuits, making LI interfaces especially expensive for fine-grained integration.

The LI wrappers provide design modularity, but there is a cost. We need additional ready and valid signals, FSMs to orchestrate them, and a FIFO for bookkeeping. These additional components result in both area (chip resources) and performance overheads and yield a more complex design that is harder to verify. Table 1 quantifies the hardware cost of LI interfaces by synthesizing designs using Vivado and comparing them to LS implementations. The LI interface requires 3–4× more registers, uses 29–31% more LUTs, and reduces the maximum frequency by 21–25%. For the higher frequency designs, the critical path for the LS implementation remains within the compute units; for the LI design, the handshaking logic itself becomes the critical path.

| Configuration | LUTs | Registers | Freq. (MHz) |
|---|---|---|---|
| **LI** (A=1, M=1) | 614 | 824 | 134.5 |
| **LS** (A=1, M=1) | 441 | 205 | 163.0 |
| **LI** (A=4, M=2) | 662 | 1426 | 224.4 |
| **LS** (A=4, M=2) | 459 | 482 | 280.8 |

**Table 1.** Resource usage of latency-sensitive (**LS**) and latency-insensitive (**LI**) FPU implementations. *A* and *M* are the latencies of the FloPoCo-generated adder and multiplier.

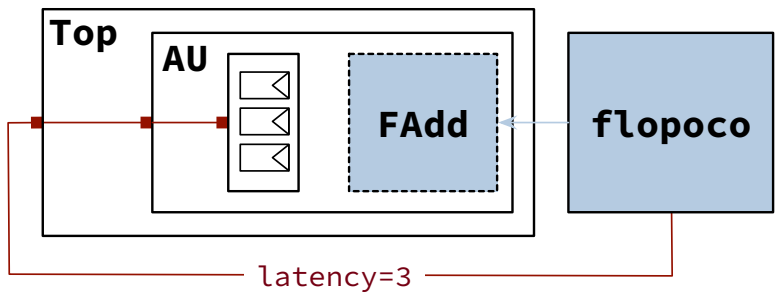

**(a)** Top-down parameterization. Adder latency is plumbed through the hierarchy to parameterize pipeline-balancing shift register.

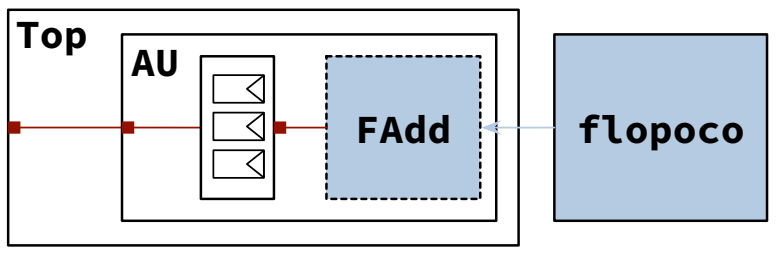

**(b)** Bottom-up parameterization. Generated modules produce output parameters resulting in clean encapsulation.

**Figure 3.** Flow of parameters when using generators.

For designs with fundamentally input-dependent timing behaviors, such as a memory hierarchy, LI interfaces are mandatory and their cost is unavoidable. However, for our FPU design, where each specific adder or multiplier has a latency-sensitive interface, the cost is pure overhead. The design-time benefit of ease of integration and correct reasoning provided by an LI interface conflicts with the need for the highest performing implementation.

### 2.3 Parameterized Design

A third alternative is to implement a *parameterized design*. Such designs use compile-time values called parameters, along with metaprogramming constructs such as loops and conditionals, to generate circuits. An HDL frontend evaluates these constructs during its *elaboration* phase. In our case, this means that if we implement an FPU using parameters for the latencies of the adder and the multiplier, we could generate efficient LS interfaces without hard-coding any timing details.

| Interface | Design | Compile | Execute |
|---|---|---|---|
| Latency Sensitive (LS) | ✓ | ✓ | ✓ |
| Latency Abstract (LA) | ✗ | ✓ | ✓ |
| Latency Insensitive (LI) | ✗ | ✗ | ✓ |

**Table 2.** When an interface's timing behavior is known.

To cope with possible timing mismatches between the adder and multiplier, our parameterized design must generate a configurable number of pipeline stages to balance the latencies. For example, if the adder takes two cycles but the multiplier takes four, the output from the adder needs to be delayed by two cycles to ensure that the FPU's latency is always four cycles. The FPU can use a parameterized shift register module, `Shift` that delays a signal by *N* cycles:

```
localparam Max = Add_L > Mul_L ? Add_L : Mul_L;
localparam Add_B = Max - Add_L;
localparam Mul_B = Max - Mul_L;
Shift#(.N(Add_B)) s_add(.in(sum), .out(sum_d));
Shift#(.N(Mul_B)) s_mul(.in(mul), .out(mul_d));
Shift#(.N(Max)) s_op(.in(op), .out(op_d));
```

This delay logic suffices for any pair of latencies for our FloPoCo-generated adder and multiplier, `Add_L` and `Mul_L`. But implementing it in current HDLs like Verilog can quickly become awkward.

The first challenge is that our FPU must somehow know these latency parameters, `Add_L` and `Mul_L`, to generate this delay logic. HDL parameters flow *top down*, from parent modules to child modules, so the only obvious solution is to add these parameters to the FPU module's signature and "plumb" them down to the delay logic (Figure 3a). But this violates encapsulation. *Users* of the FPU module should not need to know about the adder's and multiplier's latencies; FloPoCo determines the latency of the modules. Logically, we want these parameters to flow *bottom up*, from the adder and multiplier up to the FPU (Figure 3b). Current HDLs do not allow this, so users adopt various ad-hoc approaches such as using hierarchical references [19], passing them using external configuration files [23, 31], or using complex parameter negotiation frameworks [7].

The second challenge is that this parameterized approach implicitly encodes assumptions about the timing behavior of modules and the relationships between parameters. The designer of the FPU module would like to locally reason about the relationship between the parameters `Add_L` and `Mul_L`. However, an external user can easily make the mistake of flipping the two parameters when they are provided as inputs causing the FPU module to generate incorrect balancing logic. The flexibility of metaprogramming also makes it error-prone and potentially more difficult to verify.

```
gen "flopoco" comp FPAdd[#W]<G:1>( // Fully-pipelined
  val_i: interface[G]              // Provide new input
  l: [G, G+1] #W, r: [G, G+1] #W   // Required in cycle 1
) -> (o: [G+#L, G+#L+1] #W         // Provided on cycle L
) with { some #L where #L > 0; }   // Output parameter
```

**Figure 4.** Latency-abstract interface for FloPoCo adders. The latency is determined at elaboration-time.

### 2.4 Latency-Abstract Interfaces

The solution lies in recognizing that generator-produced designs occupy a unique point in the design space: their timing behaviors are parameterized at design-time but concrete at compile-time (Table 2). We name such interfaces *latency-abstract* (LA) interfaces as they abstract away timing details of modules, but do so using parameters. They capture how a module's timing behaviors are influenced by various parameters. When using a latency-abstract (LA) module, an implementation must itself be parameterized and adapt to changes in the timing behavior of the LA module. This is similar to how a module interfacing with an LI module must itself present an LI interface. The complex interdependence of parameters and timing behaviors also explains the challenges in verifying the correctness of integration of such modules.

We address these challenges by designing Lilac, the first HDL to capture LA interfaces and provide compile-time reasoning for them. Lilac uses a novel type system based on timeline types [28] which can express and statically reason about parameters' influence on a module's timing behavior and formally guarantee that there are no structural hazards or resource reuse violations in *any* parameterization of the design.

## 3 The Lilac Language

Lilac's design must solve two related problems: specifying latency-abstract (LA) interfaces of externally generated modules, and writing correct, parameterized programs using such interfaces. Lilac extends timeline types [28]—a type-system to reason about LS designs—to capture the effects of parameters on timing behaviors and introduces *output parameters* to capture the timing behaviors of generated modules. Together, these features enable it to capture interfaces from a variety of generators (§6) and provide compile-time reasoning about LA interfaces (§4).

We continue with our FPU example and show how Lilac captures design-time unknown timing behaviors using latency-abstract interfaces (§3.1), propagates output parameters to enable modular design (§3.2), and correctly adapts to changing timing behaviors (§3.3).

### 3.1 Specifying Latency-Abstract Interfaces

External generators like FloPoCo choose the timing behaviors of modules based on input parameters, such as bitwidth, as well as optimization directives. Figure 4 shows the Lilac interface for a FloPoCo-generated adder. Like a traditional parameterized module, it has an input parameter `#W` which controls the bitwidth of the ports (lines 1–4). The Lilac interface also captures the module's timing behavior with timeline types. The module defines an *event* $G$, provided by the input valid signal `val_i`, to represent when the module starts executing. The *delay* of the event (`G:1`) captures the number of cycles needed between consecutive inputs; since the delay is one, the module can accept inputs every cycle. *Availability intervals* impose constraints on inputs; the input `l` and `r` are required in the first clock cycle represented by the interval $[G, G+1)$.

The cycle in which the output is produced, i.e., the latency of the module, is determined by FloPoCo. Lilac introduces the *output parameter* `#L` to represent the latency, with a `where` clause to constrain it to be at least one. Unlike the input parameter `#W`, the output parameter is *produced* by the `FPAdd` and can be accessed by the parent module. This parameter is *abstract* during design-time, i.e., a parent module using the adder cannot assume anything about its value, except that it is at least one. During compilation (§5), Lilac's compiler executes FloPoCo to generate an adder implementation and substitute a concrete value for `#L`.

```
1  comp FPU[#W]<G:1>(
2    op: [G, G+1] 1, l:  [G, G+1] #W, r:  [G, G+2] #W
3  ) -> (o:  [G, G+1] #W) {
4    // Instantiate the modules
5    Add := new FPAdd[#W]; Mul := new FPMul[#W];
6    // Schedule execution when G occurs
7    add := Add<G>(l, r); mul := Mul<G>(l, r);
8    mx := new Mux[#W]<G>(op_s, add_s.out, mul_s.out);
9    out = mx.out; // produce output
10 }
```

**(a)** Initial, erroneous implementation.

```
comp FPU[#W]<G:1>(..) -> (o: [G+#L, G+#L+1] #W) with {
  some #L; } { .. // Instantiate and use modules.
  let #Max = Max[Add::#L, Mul::#L]::Out;
  sa := new Shift[#W, #Add_B]<'G+Add::#L>(add.out);
  sm := new Shift[#W, #Mul_B]<'G+Mul::#L>(mul.out);
  so := new Shift[#1, #Max]<'G>(op.out);
  mx := new Mux[#W]<G+#Max>(so.out, sa.out, sm.out);
  #L := #Max;   // Latency of this module
}
```

**(b)** Parameterized code to balance the pipeline.

**Figure 5.** Latency-abstract FPU implementation in Lilac. The erroneous implementation is rejected by Lilac's type system with a compile-time error. The corrected implementation balances the pipeline and itself provides a latency-abstract interface.

### 3.2 Programming with Latency-Abstract Interfaces

Section 2.3 described the challenges of correctly integrating modules whose parameters influence timing behaviors. Using LA interfaces for the adder and the multiplier, we implement an FPU (Figure 5a) and introduce Lilac programming constructs. However, this implementation is erroneous: it does not balance the pipeline when the latency of the adder and the multiplier are different and will only work when their latencies match. We demonstrate how Lilac's type-system, which builds upon timeline types [28], reasons about output parameters and ensures that the design bug is caught at compile-time.

The implementation instantiates the adder and multiplier modules (line 5) with the input parameter `#W`, which works like parameters in traditional HDLs. On line 7, it *invokes* the modules with the inputs when the event `G` occurs. Invocations (from Filament [28]) associate a particular use of an instance with when an event occurs and enable Lilac's type-system to reason about the clock cycle in which a computation occurs. When compiling this design to RTL, Lilac generates the following error:

```
mux := new Mux[#W]<G>(op, add.out, mul.out);
      Signal available in [G+Add::#L, G+Add::#L+1]
      but required in [G, G+1]
```

The error states that the output port of the adder will have a valid value on cycle `Add::#L` while the multiplexer, which is invoked on the first cycle, attempts to read its value in the first cycle. To fix the error, the user must invoke the multiplexer at the correct time. However, there is no concrete value of time at which the computation will occur; it depends upon the latency of the adder that FloPoCo generates during elaboration. To correctly schedule the computation, the user must use the *output parameter* representing the adder's latency:

```
mux := new Mux[#W]<G+Add::#L>(op, add.out, mul.out);
```

The invocation is now scheduled using the output parameter `Add::#L`: the compiler will select the exact cycle in which to execute the multiplexer after elaboration. Unfortunately, this fix is insufficient:

```
mux := new Mux[#W]<G+Add::#L>(op, add.out, mul.out);
      Signal available in [G+Mul::#L, G+Mul::#L+1]
      but required in [G+Add::#L, G+Add::#L+1]
```

While the adder's output is available in cycle `Add::#L`, the multiplier's output is not; it is available in cycle `Mul::#L` and Lilac cannot show that these are the same. A correct implementation has to balance the pipeline using the latency of *both* the adder and the multiplier. Unlike in a traditional HDL, this is a compile-time error rather than a silent error.

### 3.3 Parameterization in Lilac

```
1 comp Shift[#N]<G:1>(
2   input: [G, G+1] 32) -> (
3   out: [G+#N, G+#N+1] 32) where #N > 0 {
4   bundle<#i> w[N+1]: [G+#i, G+#i w+1] 32;
5   w[0] = input;  out = w[#N];
6   for #k in 0..#N {
7     R := new Reg[32]<G+#k>(w[#k]);
8     w[#k+1] = r.out;
9 }}
```

**(a)** Shift register implementation.

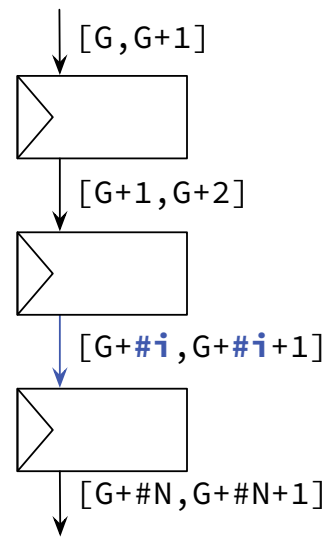

**(b)** Timing.

**Figure 6.** A shift register implementation in Lilac and a block diagram visualizing the timing of each bundle wire.

To fix the issue, we need a parameterized shift register to delay the signal. Figure 6 implements a shift register in Lilac and introduces its metaprogramming capabilities.

***Parametric signatures.*** Lines 1–3 in Figure 6a define the signature of the `Shift` module. In this case, the designer, not an external tool, controls the latency by defining an input parameter `#N` and setting the output to appear in the $N$th cycle. This means that, for example, when using the output from a `Shift[4]`, we schedule the downstream computation with `G+4` instead of using an output parameter.

***Bundles and loops.*** A Lilac program might use parameters to instantiate and schedule modules. For example, `Shift` instantiates $N$ `Reg` modules and forwards the output from register $i$ to register $i+1$. The availability interval for the $i$th register *depends* upon how many registers come before it (Figure 6b). Lilac introduces *bundles*, a multidimensional array where the availability of a value at a particular index depends on the index itself. For example, the bundle on line 4 has $N+1$ elements and states that the $i$th element's availability interval is $[G+i, G+i+1)$. The input signal is assigned to `w[0]` and the output signal is read from `w[N]` (line 5) which have availabilities $[G, G+1)$, and $[G+N, G+N+1)$ respectively. Finally, we use a compile-time for loop (lines 6–9) to instantiate registers and invoke them on the $k$th cycle, using the value from the $k$th index in the bundle as the input, and assigning the output of the register to the index $k+1$.

Like loops, bundles are a compile-time construct: after elaboration (§5), our shift register will look like the following code where the bundle has been completely eliminated and the implementation simply forwards values through a chain of registers:

```
R1 := new Reg; r0 := R0<G>(input);
R1 := new Reg; r1 := R1<G+1>(r0.out); ...
```

***Putting it together.*** The shift register lets us balance our FPU's pipeline (Figure 5b). First, we compute the maximum of the two latencies using *parameter access*. The Lilac component `Max` has an empty body and defines a single output parameter `Out` which is constrained to the expression `#A > #B ? #A : #B`; it is only used as a function over parameters. The combination of input and output parameters allows Lilac programs to use components as pure functions over parameters and abstract away routine computations. Next, we compute the amount of balancing needed (`#Add_B` and `#Mul_B`, elided), and use that to delay each compute module's outputs as well as the `op` signal. Finally, because the latency of the FPU depends on the latency of the compute module, we abstract its latency using `#L` and provide a binding in the body.

## 4 Type System

Lilac's type system analyzes each component in the program to ensure that it matches its provided signature and uses its submodules correctly (based on their signatures). To enforce that each component can be safely pipelined, Lilac uses an SMT solver to enforce that its parameterized expressions obey the requirements of timeline types [28].

### 4.1 Syntax

Figure 7 lists the syntax of the Lilac HDL, which builds upon Filament's basic structural constructs and timeline type system [28]. Filament does not support parameters. It has three main constructs: instantiations (to instantiate submodules), invocations (to schedule computations), and connections (to connect ports). Lilac adds parameterization and metaprogramming constructs.

***Parameter expressions.*** Lilac adds parameter expressions ($P$) and allows them to occur in every location where constants are allowed in Filament (availability intervals, event delays, and scheduling expressions, etc.). Parameters are defined at three locations: (1) signatures (*sig*) declare input parameters, (2) `some` bindings (*bind*) define output parameters, and (3) `let` bindings (*cmd*) name parameter expressions. The built-in arithmetic operators (*binop*) are directly encoded into SMT queries, allowing the solver to reason about computations.

More complex computations can be specified in two ways. *Parameter access expressions* allow instantiation and use of components to get their output parameters; for example, `Max[#A, #B]::#Out` instantiates the `Max` component and gets the output parameter representing the max of $A$ and $B$. Lilac also declares common operations such as $log_2$ and $exp_2$ as *uninterpreted functions* within its encoding (§4.2) and provides common equalities such as $exp_2(log_2(N)) = N$; the solver relies on these equality to prove fact and requires the user to provide addition facts using `assume` statements if more complex reasoning is required.

***Component signatures.*** Lilac signatures (*sig*) specify the events, parameters, and ports associated with a module. A signature declares output parameters using `with` bindings

$$
\begin{array}{lll}
x & \in \textit{variables} \quad p \in \textit{parameters} \quad G \in \textit{events} \\
bop & \in \{+,-,\times,\div,\%\} \quad unop \in \{\log_2, \exp_2\} \\
P & ::= \mathbb{N} \mid p \mid bop(P_1, P_2) \mid unop(P) \mid x_1[P^*]::x_2 \\
C & ::= P_1 = P_2 \mid P_1 \leq P_2 \mid \neg C \mid C \wedge C \mid C \vee C \\
ival & ::= [G_1 + P_1, G_2 + P_2] \quad port ::= x : ival\ p \\
bind & ::= \textbf{some}\ p\ \textbf{where}\ c^* \\
sig & ::= x[P^*]\langle G : P\rangle(port^*)\ \textbf{with}\ \{bind^*\}\ \textbf{where}\ C^* \\
mod & ::= \textbf{comp}\ sig\{cmd^*\} \mid \textbf{extern}\ sig \mid \textbf{gen}\ tool\ sig \\
acc & ::= x \mid x_1.x_2 \mid acc[P_1..P_2] \\
cmd & ::= cmd_1; cmd_2 \mid acc_1 = acc_2 \mid \textbf{let}\ p = P \mid p := P \\
& \mid x_1 := \textbf{new}\ x_2[P^*] \mid x_1 := x_2\langle G + P\rangle(acc^*) \\
& \mid \textbf{bundle}\langle p^*\rangle\, x[P^*] : ival^* \mid \textbf{assume}\ C \mid \textbf{assert}\ C \\
& \mid \textbf{if}\ C\{\text{cmd}_1\}\ \textbf{else}\ \{\text{cmd}_2\} \mid \textbf{for}\ p\ \textbf{in}\ P_1..P_2\{cmd\}
\end{array}
$$

**(a)** Language constructs. Filament constructs highlighted.

$$
[\![P]\!] : \mathcal{Q}_v \times 2^C \quad [\![cmd, 2^C]\!] : \mathcal{D} \times 2^C \times \mathcal{Q} \quad \text{pfunc}(x_c, x_o) : \mathcal{Q}_v
$$

$$
\text{clauses}(x_1, x_2) : C \quad \text{pargs}(x, P^*) : C \quad \text{defs}(x) : \mathcal{D} \times \mathcal{D}
$$

$$
\frac{}{[\![n]\!] = n, \emptyset} \quad \frac{}{[\![p]\!] = p, \emptyset} \quad \frac{v_1, C_1 = [\![P_1]\!] \quad v_2, C_2 = [\![P_2]\!]}{[\![(bop(P_1, P_2)]\!] = \text{bop}\ v_1\ v_2, C_1 \cup C_2}
$$

$$
\frac{C_1 = \text{clauses}(x_1, x_o) \quad C_2 = \text{pargs}(x_1, P^*) \quad f = \text{pfunc}(x_1, x_o)}{[\![x_1[P^*]::x_o]\!] = f, C_1 \cup C_2}\ \textsc{Acc}
$$

$$
\frac{\begin{array}{l} d_i, d_o = \text{defs}(x_2) \quad C_o = \bigcup_{x_o \in d_o} \text{clauses}(x_2, x_o) \\ C_p = \text{pargs}(x_2, P^*) \end{array}}{[\![x_1 := \textbf{new}\ x_2[P^*], pc]\!] = d_o, pc \cup C_o, (\textbf{assert}\ pc \Rightarrow C_p)}\ \textsc{Inst}
$$

$$
\frac{d_t, pc_t, q_t = [\![cmd_1, pc \wedge c]\!] \quad d_f, pc_f, q_f = [\![cmd_2, pc \wedge \neg c]\!]}{[\![\textbf{if}(c)\ cmd_1\ \textbf{else}\ cmd_2, pc]\!] = \emptyset, pc, (\textbf{assert}\ q_t \wedge q_f)}\ \textsc{Cond}
$$

**(b)** Selected encoding rules for generating SMT constraints.

**Figure 7.** Lilac language and rule for encoding constructs into SMT queries

and can specify constraints on them using `where` clauses. Finally, signatures can point to either Lilac modules (`comp`), modules implemented in Verilog (`extern`), or modules generated by a tool (`gen`). Lilac's compiler uses the last kind of declaration to automatically invoke the generator while elaborating the program (§5).

***Metaprogramming constructs.*** Lilac supports bounded for-loops, conditional expression, and recursive module instantiation with the standard semantics. Bundles provide a way to track availability intervals for generated signals and `assert` and `assume` allow for compile-time reasoning of facts.

### 4.2 Constraint Generation

Within a single component, there are two sources structural hazards: reading values when they are not semantically meaningful, and mapping multiple *logical* computations to the same *physical* resource in the same clock cycle. These are logical bugs: within the design, they appear as mistakes within the control logic of a circuit which might attempt to use a value on a wire before the correct amount of time has elasped, or assume that a module has accepted a input before it really does. However, such mistakes do not cause an obvious run time failure; the design will keep executing with incorrect values causing silent data corruption.

Filament [28] defines two properties to eliminate structural hazards: *latency safety* requires that values on wires are only read when they are semantically meaningful, and *resource safety* requires that, when using a partially-pipelined module, inputs are sent with appropriate delays between them.[2] Lilac enforces them by generating constraints that ensure:

1. *Valid reads:* Components only read from ports during their availability intervals, i.e., when their values are semantically valid.
2. *Non-conflicting writes:* There is only one logical driver of a port per clock cycle.
3. *Appropriate delays:* Components respect the initiation intervals of their subcomponents: if $d$ is the subcomponent's delay, then the component waits at least $d$ cycles before invoking the subcomponent again.

For each property, Lilac's type system generates a set of constraints over symbolic parameters and ensures that, no matter what parameterization of a design is selected during elaboration, it will not have such silent logical bugs.

***Encoding function.*** The encode function (Figure 7b) analyzes parameterized programs and generates constraints for the SMT solver. The encode function transforms each language construct in Lilac into a set of SMT constraints.

$$
[\![cmd, 2^C]\!] : \mathcal{D} \times 2^C \times \mathcal{Q}
$$

It takes the current command (*cmd*) and a path condition ($2^C$), which represents the set of facts true in the current branch of the parametric program, and returns a set of SMT variables to define ($\mathcal{D}$), a new path condition, and the SMT query ($\mathcal{Q}$). The rules helper functions to access the parameters defined by components (defs) and constraints on them (clauses), as well as constraints created by substituting concrete expression for input parameters (pargs).

[2] Filament calls these properties *well formedness* and *pipeline safety* respectively.

***Parameter expressions.*** Parameter expressions ($[\![P]\!]$) are encoded to generate SMT values ($\mathcal{Q}_v$)—which correspond to computations over SMT variables—and a set of constraints. Encoding numbers and parameter variables returns their SMT representation; binary operations return the equivalent SMT function, and unary functions return the application of uninterpreted functions to parameter expressions.

***Encoding output parameters.*** For each output parameter defined by a component, Lilac declares a new uninterpreted function that depends on all the input parameters of the component. For example, given the following component:

```
comp Max[#A, #B]<G:1>(..) with { some #O }
```

Lilac defines the following uninterpreted function:

```
(declare-fun Max_O ((A Int) (B Int)) Int)
```

Whenever the specific output parameter is used in the program, Lilac will replace it with `(Max_O A B)`; the Acc rule shows an example of this. This encoding allows Lilac programs to reason these specific examples:

```
FAdd[16,8]::#L == FAdd[16,8]::#L // True
Max[#A,#B]::#O == Max[#X,#Y]::#O // if #A==#X and #B==#Y
```

The instantiation rule (Inst) requires that the input expressions obey the constraints on input parameters, declares output parameters defined in the signature as new variables, and allows the path condition to assume the `where`-clause assertions on them.

***Compile-time language constructs.*** Lilac symbolically reasons about compile-time constructs such as conditionals. The Cond shows how the encoding function adds the **if** condition to the path condition when checking branches. Rules for port connections and for loops (not shown) ensure that when a signal is read, it is available according to its availability interval, and that instances are not reused more often than the component's delay allows for.

After constructing a constraint, the type system asserts its negation and asks and SMT solver [26] to find a satisfying assignment. If the solver finds such an assignment, this means that the user design violates the constraint; we can use this assignment to construct a counterexample demonstrating to the user that a set of concrete parameters values will create a bug in the design.

## 5 Elaboration

The compiler elaborates well-typed Lilac programs by executing generators to get concrete bindings for output parameters, evaluating compile-time constructs, and inlining bundles. The compiler then produces a valid Filament program which can be compiled down to a Verilog implementation. The compiler is written in 16k lines of Rust code.

| Design | Lines | Time (ms) |
|---|---|---|
| RISC 3-stage Base | 480 | 160 |
| Gaussian Blur Pyramid (§7) | 595 | 205 |
| FFT (Lilac only) | 1207 | 403 |
| FFT (using FloPoCo) | 1221 | 442 |
| Lilac's standard library | 1310 | 900 |
| BLAS Level 1 Kernels | 1346 | 1295 |

**Figure 8.** Type checker's performance.

***Top-down elaboration.*** In HDLs with only input parameters, the compiler can process modules top-down, completely processing a parent module before elaborating its submodules. This is possible because all the parameters needed to elaborate a module are already available and fully concrete. Such a scheme does not work for Lilac programs. To understand why, consider the following fragment from Figure 5b:

```
A := new FAdd[#W]; M := new FMul[#W]; // FloPoCo outputs
let #Max = Max[A::#L, M::#L]::#Out;
SA := new Shift[#W, #Max-A::#L];
```

In order to instantiate the `Shift` module, the elaborator needs concrete values for the output parameters `A::#L` and `M::#L` which will not be available until FloPoCo generates the adder and multiplier modules. Furthermore, output parameters can be used to control loops and conditionals and recursive module instantiation allows Lilac to define components with no defined order of elaboration.

***Elaboration algorithm.*** The algorithm scans the module for instantiations that only use parameter expressions that can be fully evaluated; if no such instantiation exists, then there is a cycle in the instantiation graph and the algorithm terminates with an error. The algorithm drills down the module hierarchy until it finds a completely elaboratable module which only contains references to unparameterized Lilac modules, external modules, and generator instantiated `gen` modules. Each generator provides a configuration file that defines the modules it produces and the mechanism to extract bindings for output parameters for each module (reading the command-line output, looking for a file, etc.). Using this information, the elaborator executes all loops and conditionals and instantiates `gen` modules by invoking the relevant tool. Once the generator produces a module, the elaborator collects the bindings for the module's output parameters and continues elaborating the parent module. This continues until all the modules are completely elaborated and stitched back together. After this, the elaborator runs a second pass to inline all reads and writes from bundles to get a fully structural Filament program.

### 5.1 Compiler Performance

Figure 8 summarizes the performance of Lilac's compiler. Compilation is dominated by the type checking step which generates and discharges SMT queries. Most designs take less than a second to type-check.

## 6 Latency-Abstract Interfaces in the Wild

We empirically justify our claim that LA interfaces can be used to capture interfaces for modules generated by existing tools.

### 6.1 Vivado IP Core Generators

Vivado's IP core generators provide encrypted, opaque implementations optimized for specific AMD FPGAs [18]. To study Lilac's efficacy at capturing LA interfaces, we study three IP core generators provided by Vivado.

***Multiplier.*** Like `Shift` (Figure 6), the multiplier core generator takes an explicit input parameter to specify the output latency. Lilac's LA interfaces capture this well:

```
comp Mult<G:1>[#W, #L](a: [G, G+1] #W,
  b: [G, G+1] #W) -> (o: [G+#L, G+#L+1] #W)
```

***Divider.*** The divider generator [16] provides several parameters to control the generated modules' timing behavior (Figure 9). It allows the user to select between microarchitectures based on precision and performance requirements:

`LutMult` (Figure 9a) is recommended for bitwidths less than 12, is fully pipelined, and has an eight-cycle latency. `Radix-2` (Figure 9b), recommended for bitwidths less than 16, has an input parameter (`#II`) which controls the pipelining for the module and can be set to values greater than one to reduce resource usage. Based on other input parameters, such as whether a fractional vs. integer remainder is required, it selects a particular formula to compute the module's latency (captured with output parameter `#L`). Since the signature specifies the exact formula used, the parent module can reason about the implementation's concrete latency. Finally, `High-radix` (Figure 9c) should be used for bitwidths greater than 16. The user guide provides a table that gives the implementation's exact latency based on the bitwidths of the divisor and the dividend; there is no closed form formula to compute it. The LA interface simply abstracts the latency using an output parameter.

The complexity of the divider implementation makes it challenging to integrate into existing designs. Older versions [14] only provided LS interfaces for integration but more recent versions [16] have added the ability to wrap the generated modules in an LI AXI interface. This is a routine example of the *convenience* use of LI interfaces: the dividers' timing behavior is extremely complicated to reason about and therefore uses an LI interface instead.

With Lilac, programmers avoid this unnecessary overhead by using an LA interface. Figure 9d demonstrates this approach: it provides a wrapper module that instantiates different divider implementations based on the input bitwidth and sets the appropriate output parameter values. In addition to demonstrating that LA interfaces can provide a uniform interface for the different divider implementations, this wrapper also encapsulates the documentation's guidance.

***FFT generator.*** Vivado's FFT generator [15], similar to `High-radix`, defines a table that uses the FPGA target and input parameter values to determine the module's latency and provides an option to generate an AXI wrapper. The LA interface to wrap this generator would look similar to Figure 9c.

| **Generator** | **Features** |
|---|---|
| PipelineC [21] | `in-dep` |
| FloPoCo [6] | `in-dep`, `out-dep` |
| XLS [10] | `in-dep`, `ii-gt-1` |
| Spiral FFT [25] | `in-dep`, `out-dep`, `ii-gt-1` |
| Aetherling [8] | `in-dep`, `out-dep`, `ii-gt-1`, `multi` |

**Table 3.** Generators integrated with Lilac and features needed to capture their interfaces. `in-dep`: input parameters affect timing behaviors, `out-dep`: output parameters affect timing behaviors, `ii-gt-1`: parameter-dependent pipelining, and `multi`: requires multi-cycle intervals.

### 6.2 LA Interfaces for Generators

Table 3 summarizes existing generators and the Lilac features needed to capture their interfaces.

- **Input parameter dependent timing (`in-dep`).** Similar to Vivado's multiplier core generator, PipelineC allows user to specify the exact latency they want using an input parameter.
- **Output parameter dependent timing (`out-dep`).** FloPoCo uses both input parameters and high-level optimization goals to decide the timing behaviors of a module. Similar to the `High-Radix` divider, such modules need Lilac's output parameters to capture their design-time abstract timing behaviors.
- **Parameter dependent pipelining (`ii-gt-1`).** XLS can generate modules that are partially pipelined, i.e., have an initiation interval of greater than one. The pipelining can depend on input parameters (similar to `Radix-2`) or be completely abstract, requiring output parameters and Lilac's capability to capture parameter-dependent event delays.
- **Multi-cycle interval (`multi`).** Aetherling generates modules where the user must hold the input signal stable for more than one cycle. Capturing such interfaces requires Lilac's ability to describe multi-cycle availability intervals.

```
comp LutMult<G:1>[#W](
  n: [G, G+1] #W,
  d: [G, G+1] #W
) -> (
  q: [G+8, G+9] #W
)
```

(a) `LutMult` architecture, latency-sensitive.

```
comp Rad2<G:II>[#W,#II,#Fr](
  n: [G, G+1] #W, d: [G, G+1] #W
) -> (q: [G+#L, G+#L+1] #W) with {
  some #L } where #II < 9, #II%2,
  #Fr & #II > 1  ? L == #W+5
: #Fr & #II == 1 ? L == #W+4 ..
```

(b) `Radix-2` architecture, input parameter dependent timing.

```
comp HighRad<G:#L>[#W](
  n: [G, G+1] #W,
  d: [G, G+1] #W
) -> (
  q: [G+L, G+L+1] #W
) with { some #L; }
```

(c) `High-radix` architecture, latency-abstract.

```
if #W<12 { D := new LutMult[..];
  #L := 8; #II := 1 }
else if #W<16 { D := new Rad2[..];
  #L := D::L; #II := D::II; }
else { D := new HighRad[..];
  #L := D::L; #II := D::L; }}
```

(d) Selecting dividers based on input bitwidth.

**Figure 9.** Interfaces for dividers produced by Vivado's IP Core. The wrapper code provides a uniform selects the implementation based on input bitwidth and provides a stable interface to integrate any divider.

Lilac allows integration of external Verilog modules by defining *external* components which provide a Lilac signature as well as a path to the Verilog file that contains the implementation. During compilation, the Lilac compiler links-in the implementation of the module with the rest of the design.

## 7 Gaussian Blur Pyramid

Gaussian blur pyramid is multi-scale image processing algorithm commonly used in tasks such as object detection. We use GBP to compare LA and LI interfaces. We implement two versions that use convolution modules generated by Aetherling [8]: one uses Lilac and LA interfaces; the second uses Verilog and LI interfaces. We qualitatively evaluate both interfaces for their ability to effectively encapsulate the generated modules and report on the resource and performance overhead incurred by LI implementation.

### 7.1 Implementation

The algorithm proceeds by repeatedly applying Gaussian blur by convolving the image with a filter, downsampling the resulting image, and repeating this to construct different *levels*. Once all levels are created, the algorithm upsamples the images, blurs them to remove aliasing artifacts, and blends them between level $N-1$ and $N$ by taking a weighted average between the two, continuing until it has recovered the original image dimensions. Our GBP implementations unroll this dataflow graph and instantiate modules to perform the blurring, blending, and up- and downsampling.

***Convolution interface.*** We use Aetherling [8], a DSL for generating stream processing programs, to generate $4 \times 4$ convolution implementations and vary the number of multipliers used to express area–performance trade-offs. We design a Lilac-based LA interface (Figure 10a) and a Verilog-based LI interface (Figure 10b) and analyze how they capture the timing behavior of the generated modules.

***Number of inputs.*** First, upon varying the number of multipliers, Aetherling changes the size of the input port to accept $N$ elements at a time, where $N$ is a factor of 16. Lilac defines an output parameter to express this tool-dependent

```
comp AethConv[#W]<G:#II>( // 4x4 conv, gauss filter
  valid_i: interface['G],
  in[#N]: ['G, 'G+#H] #W  // Num. inputs & hold time
) -> (out[#N]: ['G+#L, 'G+#L+1]) with {
  some #H where #H > 0;   // Num. cycles to hold inputs
  some #N where 16 % #N == 0, #N > 0; // Chunk size
  some #L, #II where #L > 0, #II >= #H }
```

(a) Latency-abstract interface for convolution in Lilac.

```
module AethConv #(parameter W=32, N=16, L=6, II=1, H=1)(
 input logic val_i, input logic rdy_i,    // input intf
 input logic [N-1:0][W-1:0] in,
 output logic val_o, output logic rdy_o,  // output intf
 output logic [N-1:0][W-1:0] out);
```

(b) Latency-insensitive interface for convolution in Verilog.

**Figure 10.** Interfaces for the Aetherling-generated $4\times 4$ convolution module. The parameter $N$ is used by the implementation to describe how many of the 16 required inputs it will accept in a cycle and partially streams some outputs.

choice. One the other hand, the LI interface defines an input parameter that *must be* correctly threaded through the hierarchy; if the user gets it wrong, the implementation may silently ignore the inputs.

***Timing behaviors.*** The LA interface defines three output parameters to capture the module's timing behavior. The latency (`#L`) and initiation interval (`#II`) have corresponding signals in the LI interface, namely an output valid signal (`val_o`) and a ready-in (`rdy_i`) signal. Aetherling's partially-pipelined modules use multipliers that perform computations over multiple cycles and require the module to hold the input signal stable for multiple cycles. The third parameter, `#H`, captures this. The LI interface lacks an explicit signal to capture this information: it assumes that data transfer occurs when ready and valid are asserted together. Instead, we plumb the `#H` parameter through the hierarchy and use it to latch the input value for the required number of cycles.

***Parameter pollution.*** The LI design suffers from *parameter pollution:* the top-level GBP module, which uses three separate blur modules, must pass four parameters to each,

```
1 comp Ser[#W, #N, #B, #C, #H]<G: #C*(#N-1)+#H>(
2   en: interface[G], in[N*B]: [G, G+#H] W
3 ) -> (o[#N][#B]: [G+#C*#i, G+#C*#i+#H] W) {
4   for #i in 0..#N { for #j in 0..#B {
5     let #CurIdx = #B*#i+#j;
6       d := new Reg[#W]<G, G+C*#i+#H>(in{#CurIdx});
7       o{i}{j} = d.out; }}}
```

**Figure 11.** Parameterized, pipelined serializer in Lilac. The interface specifies how to take $N \times B$ elements of an array and chunk them into a stream of $N$ element bundles and can control gap between consecutive bundles ($C$) and the number of cycles a bundle is available ($H$).

```
1  case (st)  // send state machine
2    IDLE: if (val_i) nxt_st = PROC;
3    PROC: cv_val_i = 1; if (cv_rdy_i) nxt_idx = idx+1;
4    BLOCKED: if (rdy_o) nxt_st = IDLE;
5  assign conv_in = in[N*idx+:N];
6  case (st) // recv state machine
7    IDLE: if (val_i) nxt_st = PROC;
8    PROC: cv_rdy_o = 1; if (cv_val_o) nxt_idx = idx+1;
9    BLOCKED: if (rdy_o) nxt_st = IDLE;
10 assign out[N*idx+:N] = conv_out;
```

**Figure 12.** State machines for pipelined execution of Aetherling-generated convolution in the LI implementation.

resulting in 12 extra parameters flowing through the design. For larger GBPs, this quickly becomes infeasible.

***LA blur implementation.*** Different levels of the GBP algorithm apply blurs of different sizes. For both implementations, we provide a parameterized blur module that takes an image of size $I \times I$ and repeatedly calls the Aetherling generated convolution module to process it. For the LA implementation, this is accomplished using a serializer module (Figure 11) which takes $I^2$ elements from the image and provides $N$ inputs to the convolution module (where $N$ is the output parameter provided by the Aetherling module). The serializer interface (lines 1–3) provides control over when bundles are produced, if there are gaps between them, and how long they are valid for. However, the module implementation (lines 4–7) simply instantiates a register for each element and forwards its output; Lilac's type system ensures that the complex reasoning about parameter-dependent bundle availability is correctly handled.

***LI blur implementation.*** The LI blur implementation uses two state machines (Figure 12) to enable pipelined execution of the convolution module. The `send` state machine (lines 1–5) repeatedly extracts and sends an $N$-sized chunk of data to the Aetherling module; the `cv_rdy_i` indicates when the convolution is ready to accept the inputs. The `recv` state machine (lines 6–10) waits for the output (`cv_val_o`) and stores it in the right location in the output image.

| Design (N) | LUTs | Registers | Freq. (MHz) |
|---|---|---|---|
| Lilac / RV (1) | 1824 / 2093 | 2532 / 3254 | 258 / 236 |
| Lilac / RV (2) | 1762 / 2062 | 2464 / 3165 | 284 / 219 |
| Lilac / RV (4) | 1627 / 1983 | 2373 / 3129 | 270 / 306 |
| Lilac / RV (8) | 1227 / 2146 | 1733 / 3058 | 223 / 231 |
| Lilac / RV (16) | 1311 / 2099 | 1688 / 3244 | 211 / 183 |

**Figure 13.** Resource usage and maximum frequency of GBP implementations for different convolutions configurations.

***Pyramid implementation.*** For the LA implementation, the various uses of the Blur module require pipeline balancing. Lilac's type system ensures that we correctly delay and use the signals when passing them to subsequent stages:

```
Blur0 := new BlurAbs[8]; Blur0 := new Blur[4];
level0 := Blur0<'G>(input);
down := new Down<G+Blur0::#L>(level0.out);
level1 := Blur1<'G+#Blur0::#L>(down.out); ...
#L := Blur0::#L + Blur1::#L + BlurUp::#L + Blend::#L;
// II is dictated by slowest blur.
#II := Max[Blur0::#II, Blur1::#II, BlurUp::#II]::#Out;
```

The LI implementation uses a serial state machine to execute each blur module through a ready–valid interface and itself provides a ready–valid interface.

### 7.2 Cost of Latency Insensitivity

Figure 13 summarizes our results: in general, LI implementations of the GBP take more resources to achieve the same maximum frequencies. We implement five design points for both the Lilac and the latency-insensitive interface by synthesizing Aetherling modules with different values for parameter $N$, which affects the latency and the throughput of the design. Each design is synthesized using Vivado v2023.2 with a target clock period of 3ns and input-output delays of 0.1ns. The reported frequency is computed by subtracting the worst negative slack from the target and we take geometric means when reporting summary statistics.

***Performance analysis.*** On average, the LI designs achieve 6.8% worse frequencies, and use 26.2% more LUTs and 33.0% more registers compared to the Lilac implementation. As expected, the resource usage of the LI designs remains roughly constant across the various design points. However, for the LA implementations, the resource usage *goes down* as we increase the value of $N$: Lilac-1 uses 22.2% fewer registers than RV-1 while Lilac-16 uses 48% fewer registers. This is because the serialization logic is a primary cost of the LA implementation and, as the convolution module provides more parallelism, it needs less serialization logic.

The trend illustrates an important trade-off: LI interfaces requires a large upfront resource cost to implement the handshaking FSM; however, this cost remains constant regardless of the internal logic of the module. LA interface, on the other hand, impose a parameter-dependent cost because of their reliance on compile-time code-generation. This points of a design paradigm where designers connect smaller components using LA interfaces and, once they have a big enough "island" of LA modules, wrap them using an LI interface.[3]

***Design analysis.*** For the LI designs, the state machine logic within the blur modules is the critical path; for the Lilac implementation, it is the serializer implementation due to high fanout. The resource overheads in the LI modules comes from the extra state machines and are amortized as the size of the module increases implying that designers should prefer using LA interfaces to build large components which can be wrapped using LI interfaces.

## 8 Related Work

***Parameters in HDLs.*** Traditional HDLs, like Verilog and VHDL, support parameterizing designs using input parameters and metaprogramming constructs. They also provide mechanisms such as `defparam` to assign parameter values anywhere in a design's hierarchy; however, such constructs break encapsulation and are discouraged from use due to poor tool support [17, 19]. *Output parameters* capture inverted parameter flow, provide strong encapsulation, and work with Lilac's type system to ensure correct usage. Embedded HDLs [2, 3, 20, 23] do not use parameters in the traditional sense: they rely on the host software language to construct circuits and combine them. Because of this, eHDLs can easily implement output parameters. However, unlike Lilac, they do not capture the influence of parameters on timing behaviors and therefore make LA interfaces challenging to use correctly.

***Safe HDLs.*** Traditional HDLs guarantee *datatype safety*: they ensure that the bits on every wire correspond to semantically meaningful data. Safe HDLs provide orthogonal guarantees about how circuit elements are used. Latency-counting HDLs [2, 13, 32, 33] track the latencies of signals and ensure that pipelines are correctly balanced (*latency safety*) but cannot reason about partially-pipelined modules. Filament [28] allows designs to be partially pipelined and reasons about how resources such as wires, registers, and modules are reused over time, either explicitly or through pipelining, and statically guarantees the absence of resource reuse bugs (*resource safety*). By enforcing both latency safety and resource safety, Filament is the first language to eliminate all structural hazards at compile time. Anvil [35] is a channel-based HDL that adapts Filament's abstractions of events and availability intervals and adds support for LI interfaces using synchronization signals. However, Anvil does not support explicit resource reuse, does not address the interaction between compile-time parameters and interface behavior, and does not aim to express LA interfaces. Lilac is the first language to statically provide both latency safety and resource safety for parameterized designs directly support LA interfaces.

***Pre-elaboration parameter checking.*** Bluespec [29] reasons about parameter values at compile-time using *provisos*, which are similar to Lilac's `where` clauses. However, Bluespec does not capture modules' timing behaviors and thus cannot reason about the impact parameters have on them.

***Verification systems.*** Formal verification tools for hardware design [4, 24, 34] can reason about general temporal properties but generally require whole program proofs and cannot directly handle parameterized designs. Lilac, in contrast, uses a compositional type system to guarantee the absence structural hazards and resource-reuse violations and reasons symbolically about parameterized designs.

***Hardware generators.*** Core generators [6, 18, 25], high-level synthesis [30, 36], and accelerator design languages [8, 10, 11, 22, 27] operate with higher-level, computational abstractions and do not expose timing concerns to the user. In contrast, Lilac is an HDL that directly expresses and reasons about timing behaviors of low-level circuit descriptions.

## 9 Future Work

Latency-abstract (LA) interfaces are a design abstraction that provide the flexibility of latency-insensitive (LI) interfaces as well as the efficiency of latency-sensitive interfaces. They do not replace LI interfaces: a module with input-dependent timing behaviors, such as a variable-latency divider, must use an LI interface. Similarly, there are cases where even though LA interfaces could work, an LI interface might be a better choice because of the design's performance objectives. These trade-offs imply the need for new HDLs that unify LI and LA abstractions and allow designers to seamlessly move between the two which we see as a fruitful area for future work.

## 10 Conclusion

A combination of generators, domain-specific languages, and HDLs allows modern hardware designs to be *malleable* and adapt to changes in performance goals and implementation strategies. Latency-abstract interfaces, coupled with safe abstractions of Lilac, allow designers to build such malleable designs without having to pick between performance, composition, and correctness.

[3] This mirrors the concept of globally-asynchronous, locally-synchronous design where large islands of synchronously clocked modules are connected using asychronous clock-domain crossings.

## Acknowledgments

We thank Christopher Batten for insightful discussions and early feedback as well as anonymous reviewers for helping clarify the framing of LA interfaces as a design abstraction. This material is based upon work supported by the NSF under Awards No. F569084 and 2237379. The first author was partly supported by the Jane Street graduate research fellowship.